\documentclass{article}

\usepackage{spconf,amsmath,graphicx}
\usepackage[hidelinks]{hyperref}
\usepackage{amssymb}   

\usepackage{booktabs} 
\usepackage{adjustbox}
\usepackage[table]{xcolor}

\title{StereoFoley: Object-Aware Stereo Audio Generation from Video}
\name{
\begin{tabular}{c}
Tornike Karchkhadze$^{1}$\sthanks{Work completed during internship at Apple.}, 
Kuan-Lin Chen$^{2}$, Mojtaba Heydari$^{2}$, Robert Henzel$^{2}$, \\
Alessandro Toso$^{2}$, Mehrez Souden$^{2}$, Joshua Atkins$^{2}$
\end{tabular}
}

\address{$^{1}$UC San Diego 
$^{2}$Apple
}

\begin{document}
\ninept
\maketitle
\begin{abstract}
We present StereoFoley, a video-to-audio generation framework that produces semantically aligned, temporally synchronized, and spatially accurate stereo sound at 48 kHz. While recent generative video-to-audio models achieve strong semantic and temporal fidelity, they largely remain limited to mono or fail to deliver object-aware stereo imaging, constrained by the lack of professionally mixed, spatially accurate video-to-audio datasets. 
First, we develop a base model that generates stereo audio from video, achieving performance on par with state-of-the-art V2A models in both semantic accuracy and synchronization.
Next, to overcome dataset limitations, we introduce a synthetic data generation pipeline that combines video analysis, object tracking, and audio synthesis with dynamic panning and distance-based loudness controls, enabling spatially accurate object-aware sound. Finally, we fine-tune the base model on this synthetic dataset, yielding clear object–audio correspondence. 
Since no established metrics exist, we introduce a stereo object-awareness metric and report it alongside a human listening study; the two evaluations exhibit consistent trends. 
This work establishes the first end-to-end framework for stereo object-aware video-to-audio generation, addressing a critical gap in the field.

\end{abstract}
\section{Introduction}
\label{sec:intro}

Recent advances in audio generation~\cite{pmlrv202liu23f,evans2025stable} have demonstrated the ability to synthesize plausible sounds with different conditioning modalities. Our prior work, ImmerseDiffusion~\cite{heydari2025immersediffusion}, we showed diffusion models can generate spatial audio with user-controlled localization. Video-to-Audio (V2A) generative models have achieved strong semantic alignment and temporal synchronization ~\cite{Iashin2021,wang2024frieren,zhang2024foleycrafter,xing2024seeing,viertola2024temporally,Cheng2025mmaudio}, yet they largely remain restricted to mono audio. Recent efforts explore stereo rendering, albeit via stereo codecs~\cite{tian2025audiox,liu2025think} or post-hoc mono-to-stereo upmixing~\cite{wang2025klingfoley}, rather than grounding spatialization in object positions. Other efforts move beyond mono to first-order ambisonics—either extracting 2D videos and explicit camera-direction cues from $360^\circ$ videos~\cite{kimvisage} or directly using panoramic videos~\cite{Liu2025OmniAudio}. A separate line targets 5.1 surround sound~\cite{dagli2024see2sound}, but models scene-level sound fields rather than object-synchronous sound generation. To the best of our knowledge, no prior work directly addresses object-aware stereo sound generation—V2A generation for 2D videos, where sounds emanate from the positions of their corresponding video objects in stereo sound field. We believe this gap exists partly due to the limitations of widely used datasets for V2A such as VGGSound~\cite{chen2020vggsound} and AudioSet~\cite{gemmeke2017audio}, which lack spatially reliable stereo audio content.

\begin{figure}[t]
  \centering
  \includegraphics[width=1.0\linewidth]{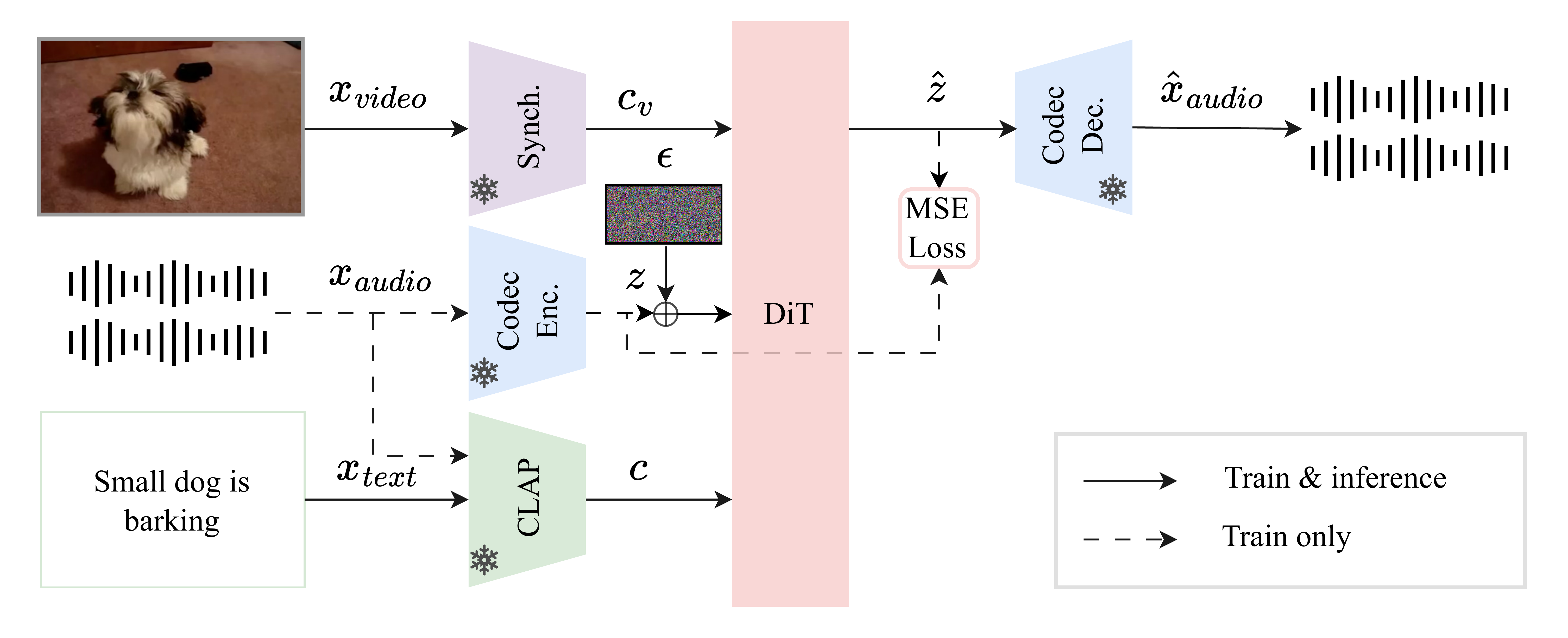}
  \caption{StereoFoley system overview: video, audio, and text encoders with a Diffusion-Transformer backbone.}
  \label{system_diagram}
  \vspace{-0.6cm}
\end{figure}

\begin{figure*}[t]
  \centering
  \includegraphics[width=1.00\linewidth]{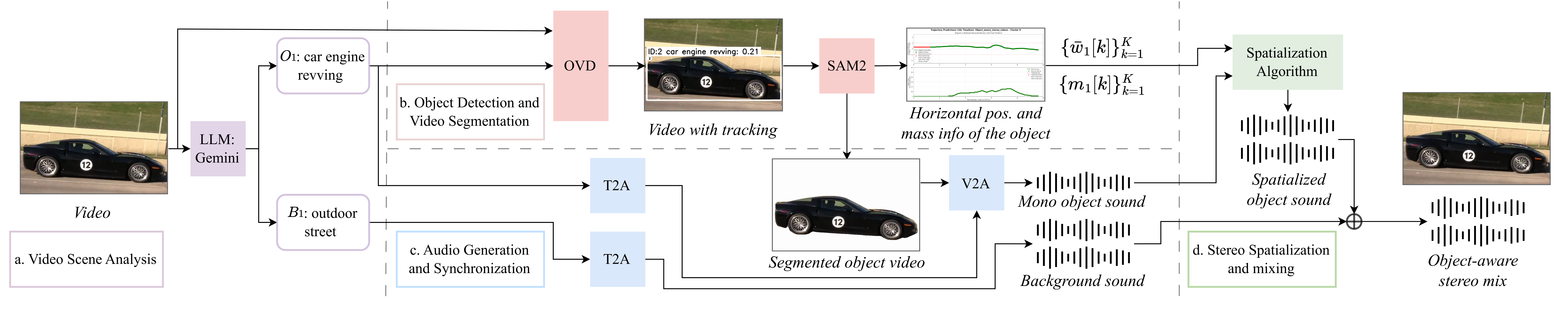}
  \caption{Overview of the object-aware stereo data generation pipeline. (a) video scene analysis with LLM, (b) object detection and tracking with segmentation, (c) audio generation and synchronization with T2A and V2A models, and (d) stereo spatialization via dynamic panning and distance-based loudness, mixing with generated background sound, producing the final object-aware stereo mix for the video.}
  \label{fig:synth_data_pipeline}
  \vspace{-0.5cm}
\end{figure*}

To address this gap, we present StereoFoley, a V2A framework that generates semantically aligned, temporally synchronized, and spatially consistent stereo audio. We first build and train a base model for stereo audio generation conditioned on video (StereoFoley-base) using VGGSound and AudioSet datasets, achieving state-of-the-art performance.
To add video object awareness, we introduce a synthetic data pipeline that adopts standard cinematic conventions of stereo rendering~\cite{holman2010sound}, where objects are panned according to their on-screen position using amplitude panning, objects remain audible off-screen to ensure permanence, and background ambiance is spread across both channels. 
In this pipeline, we track and segment objects in the video. We then generate audio with a text-to-audio (T2A) model and synchronize it with segmented video using base model. Finally, the sounds are spatialized via dynamic panning and distance-based loudness, while background ambience is synthesized and mixed in separately.
Fine-tuning our base model on this dataset yields StereoFoley-obj, an object-aware stereo model that produces audio aligned with visual object positions.

We compare our approach with current state-of-the-art MMAudio~\cite{Cheng2025mmaudio} and Kling-Foley~\cite{wang2025klingfoley}, and show that our method achieves comparable or better performance across objective metrics. Additionally, we introduce an objective measure of V2A object-aware stereo spatialization, and conduct a user study. Both evaluation methods show improved quality in stereo consistency.

\section{Method}
\label{sec:method}

\subsection{Model Architecture}

Fig.~\ref{system_diagram} shows the architecture of StereoFoley, which is based on latent diffusion~\cite{pmlrv202liu23f} and consists of two main components: encoders for video, audio, and text, and a generative diffusion base. 

Let us denote the inputs of the model as: 
audio $x_{\text{audio}} \in \mathbb{R}^{(T  \times f_s) \times 2}$ with sampling rate $f_s$ and stereo channels; 
text $x_{\text{text}} \in \mathbb{V}^L$, a sequence of $L$ tokens from the vocabulary $\mathbb{V}$; 
and video $x_{\text{video}} \in \mathbb{R}^{(T  \times f_v) \times H \times W \times 3}$ with frame rate $f_v$, video height $H$, width $W$, and $3$ RGB channels. 
Here $T$ is a video clip duration.

\subsubsection{Encoders}

We encode audio using an in-house stereo codec inspired by~\cite{kumar2023high, evans2025stable}, which maps $x_{\text{audio}}$ into latent representations $z \in \mathbb{R}^{T_z \times D_z}$, where $T_z$ and $D_z$ denote the temporal and embedding latent dimensions, respectively.
$x_{\text{text}}$  and $x_{\text{audio}}$  are mapped by in-house CLAP~\cite{wu2023large, elizalde2024natural} model into a shared embedding $c \in \mathbb{R}^{D_c}$ used for conditioning. 

For \( x_{\text{video}}\), following the success demonstrated in MMAudio~\cite{Cheng2025mmaudio}, we employ Synchformer~\cite{iashin2024synchformer}, which produces video embeddings $c_v \in \mathbb{R}^{T_v \times D_v}$, with $T_v$ and $D_v$ temporal and embedding dimensions.
Unlike MMAudio, which augments synchronization through aligned RoPE embeddings~\cite{su2024roformer} and ConvMLPs, we adopt a more simple and direct strategy: in our designed temporal resolution of audio and video latents match $T_v = T_z$, enabling video and audio embeddings direct integration, achieving strong temporal alignment without additional architectural overhead.

\subsubsection{Generative Base}

We employ a Diffusion-Transformer (DiT)~\cite{peebles2023scalable} as our generative base, composed of cascaded blocks that include self-attention, cross-attention, and gated MLP components, together with layer normalization and skip connections, following a similar architecture to~\cite{evans2024long}. The model accepts multiple conditioning signals: text and audio embeddings from CLAP, and video embeddings from Synchformer. Projection layers are used to match the feature dimensionality requirements of each conditioning pathway. CLAP features are provided through cross-attention layers, while video conditioning is directly added to the incoming noisy audio latent. The model is trained using the $v$-objective approach~\cite{salimans2022progressive}, minimizing the Mean Squared Error (MSE) between the ground-truth velocity $v$ and the predicted velocity $v_\theta(z_t, t, \{c, c_v\})$ at timestep $t$, conditioned on $c$ and $c_v$ given the latent variable $z_t$:

\begin{equation}
\mathcal{L}_{\text{Diffusion}} = \left\| v_\theta(z_t, t, \{c, c_v\}) - v \right\|_2^2
\end{equation}

\subsection{Synthetic data generation pipeline}
\label{sec:synthdatapipeline}

To address the lack of professionally mixed stereo V2A data and generate spatially accurate dataset, we build a synthetic pipeline, outlined in Fig.~\ref{fig:synth_data_pipeline} and following subsections with stages (a–d). 

\subsubsection{Video Scene Analysis (Fig.~\ref{fig:synth_data_pipeline}.a)}

We begin by generating detailed captions for videos using the multimodal large language model (LLM) Gemini 2.5-Flash~\cite{comanici2025gemini25pushingfrontier}. The model is prompted with enumerated video frames and a task description to produce high-level textual summaries of visible sound-making objects and background context. The model analyzes the video in a scene-based manner, with each scene represented as $s_i = \{T_i, O_{i,j}, B_i\}$, where $T_i$ denotes the scene duration with start and end times, $O_{i,j}$ is the set of sentences describing visible sound-making objects, $B_i$ is the background description of scene $s_i$, and $i$ and $j$ index scenes and objects, respectively.

\subsubsection{Object Detection and Video Segmentation (Fig.~\ref{fig:synth_data_pipeline}.b)}
We use LLM scene descriptions to guide object detection. We employ an in-house open-vocabulary detection (OVD) model, inspired by~\cite{ChengS24yolo}, which detects objects from free-text queries and produces bounding boxes. Before feeding sentences into the OVD model, we parse them into nouns and verbs and filter out generic plurals (e.g., “people,” “cars”) to focus on single, trackable objects. The detected boxes are then grouped into visual clusters $O_{i,j} \rightarrow c_{i,j}$, where nouns from the same sentence are grouped considering spatial proximity and scene boundaries. 
For instance, for a sentence “person playing guitar,” detections for “person” and “guitar” are merges into a single cluster by matching the “guitar” detections to the nearest “person” detections, treating them as one sounding object rather than two.
Then, we calculate pixel-level masks and segment clusters from the videos using the SAM2~\cite{Ravi2025sam2} video segmentation model. For each cluster $c_{i,j}$ in scene $s_i$, we compute its frame-by-frame pixel masses and normalized geometric center of masses positions:
\begin{equation}
\resizebox{0.48\textwidth}{!}{$
\displaystyle
m_{i,j}[k] = \lVert \Omega_{i,j,k} \rVert,\:
(\bar w_{i,j}[k],\bar h_{i,j}[k]) = \frac{1}{\lVert \Omega_{i,j,k}\rVert}
\sum_{(w,h)\in\Omega_{i,j,k}}
\left(\tfrac{2w}{W}-1,\,\tfrac{2h}{H}-1\right),
$}
\label{mass_and_centers}
\end{equation}
where $\Omega_{i,j,k}$ denotes the set of pixels assigned to cluster $c_{i,j}$ in frame $k$, and $(w,h)$ are pixel coordinates. The sequence $\{(\bar w_{i,j}[k],\bar h_{i,j}[k]), m_{i,j}[k]\}_{k=1}^{K_i}$ therefore provides the cluster’s centroid trajectory normalized to $[-1,1]$, and its per-frame pixel masses, where $K_i = T_i \times f_v$ is the number of frames in scene $s_i$.

\subsubsection{Audio Generation and Synchronization (Fig.~\ref{fig:synth_data_pipeline}.c)}
For each cluster $c_{i,j}$, generic object-specific sounds are first generated using our in-house text-to-audio (T2A) model conditioned on the LLM-derived descriptions $O_{i,j}$. These preliminary sounds are then temporally aligned with the corresponding cluster using our pre-trained StereoFoley-base model, which takes as inputs the segmented video and the generic audio encoded via CLAP to regenerate synchronized audio $a_{i,j}$ that matches the visual stream. 
We found this two-stage strategy more stable than directly conditioning V2A on segmented video, which often produced inconsistent or hallucinated outputs. 
Background (non-object) sounds are generated separately using the same T2A model conditioned on $B_i$ and mixed without temporal alignment, since they are not tied to specific visual events.

\subsubsection{Stereo Spatialization and Mixing (Fig.~\ref{fig:synth_data_pipeline}.d)}

Given cluster's horizontal centroid trajectory $\{(\bar w_{i,j}[k]\}_{k=1}^{K_i}$ and pixel masses $\{m_{i,j}[k]\}_{k=1}^{K_i}$, we spatialize the generated audio $a_{i,j}$ using standard cinematic stereo conventions for 2D video, mapping horizontal motion to audio panning and mass to loudness.

To ensure stability and reduce jitter and abrupt jumps between frames, we smooth trajectories and masses data using weighted moving average. Missing detections are filled by linear interpolation between the last and next reliable frames, while off-screen or truncated segments are extrapolated using the average recent velocity. Pixel masses are similarly smoothed, with persistence near screen edges to avoid implausible shrinkage when only part of the object is visible. For off-screen periods, we approximate the mass with the last reliable value (or the first upon re-entry). Finally, horizontal positions and masses are up-sampled from the video frame rate to the audio sampling grid. For a scene $s_i$ of duration $T_i$, this yields $N_i= T_i \times f_s$ frames of $\{\bar w_{i,j}[n], m_{i,j}[n]\}_{n=0}^{N_{i}-1}$. 

We clip horizontal positions $\hat w_{i,j}[n] = \Pi_{[-1,1]} \big(\bar w_{i,j}[n]\big)$ and then computed equal-power panning gains as:
\begin{equation}
\resizebox{0.43\textwidth}{!}{%
$g^L_{i,j}[n] = \cos(\tfrac{\pi}{4}(\hat w_{i,j}[n]+1)),\qquad
g^R_{i,j}[n] = \sin(\tfrac{\pi}{4}(\hat w_{i,j}[n]+1)).$%
}
\end{equation}
We model how distance to an object—approximated by its pixel mass—affects sound using a normalized mass term, with additional attenuation only when the object is off-screen:
\begin{equation}
\scalebox{1.15}{$
v_{i,j}[n] = 
\frac{m_{i,j}[n]}{m_{\max}}
\cdot \frac{1}{1+\max(0,|\bar w_{i,j}[n]|-1)^2},
$}
\end{equation}
where $m_{\max}=\max_k (\tilde m_{i,j}[n])$ is the maximum observed size of the object in the scene.  Thus, the mono audio $a_{i,j}[n]$ of given cluster $c_{i,j}$ is rendered into stereo as:  
\begin{equation}
\resizebox{0.44\textwidth}{!}{$
s^L_{i,j}[n] = v_{i,j}[n]\,g^L_{i,j}[n]\,a_{i,j}[n], \qquad
s^R_{i,j}[n] = v_{i,j}[n]\,g^R_{i,j}[n]\,a_{i,j}[n].
$}
\end{equation}

Finally, the spatialized cluster audio streams are mixed with generated background ambience to produce the final cinematic style stereo sound, which is subsequently used to fine-tune our model for object-aware stereo generation.

\section{Experiments and Results}


\begin{table*}[h]
\centering
\caption{ Comparison of baselines and our model across metrics of audio quality, diversity, semantic aliment and synchronization.}
\label{tab:quant_comparison}
\begin{adjustbox}{width=0.95\textwidth}
\begin{tabular}{@{}lcccccccc|cc@{}}
\toprule
\textbf{Method}  & \textbf{FD$_\text{PaSST}$$\downarrow$} & \textbf{FD$_\text{PANNs}$$\downarrow$} & \textbf{FD$_\text{VGG}$$\downarrow$} & \textbf{KL$_\text{PANNs}$$\downarrow$} & \textbf{KL$_\text{PaSST}$$\downarrow$} & \textbf{IS$\uparrow$} & \textbf{IB-score$\uparrow$} & \textbf{DeSync$\downarrow$} 
& \textbf{Stereo-Score$\uparrow$} \\
\midrule
MMAudio~\cite{Cheng2025mmaudio}  & \textbf{60.60} & \textbf{4.72} & \textbf{0.97} & 1.65           & 1.40          & 17.40 & \textbf{33.22} & 0.44 & -- \\
Kling-Foley~\cite{wang2025klingfoley} & -- & 7.60 & -- & 1.86 & -- & -- & 30.75 & 0.43 & -- \\

\midrule

StereoFoley-base (vgg only) & 64.55 & 7.21          & 1.47 & 1.64 & 1.37 & 20.15 & 30.61 & 0.42 &
0.21  \\

StereoFoley-base (vgg+Audioset)          & 62.57 & 7.83 & 1.45 & \textbf{1.63} & \textbf{1.31} & \textbf{20.36} & 31.55 & \textbf{0.41} & 0.21 \\ 

\midrule

StereoFoley-obj (vgg only) & 74.00 & 7.77 & 1.41 & 1.74 & 1.46 & 18.49 & 29.23 & 0.43 & \textbf{0.24} \\

\bottomrule
\end{tabular}
\end{adjustbox}
\vspace{-0.3cm}
\end{table*}

\begin{table}[t]
\centering
\footnotesize
\caption{Objective and subjective stereo alignment results.}
\label{tab:alignment}
\begin{adjustbox}{width=0.45\textwidth}
\begin{tabular}{lccc}
\toprule
\multicolumn{4}{c}{\textbf{Objective Evaluation (BAS)}} \\
\midrule
& on-screen & off-screen & Combined \\
\midrule
MMAudio          & 0.07 & 0.01 & 0.08 \\
VGGSound original     & 0.23 & 0.20 & 0.23 \\
StereoFoley-base & 0.23 & 0.21 & 0.23 \\
StereoFoley-obj  & \textbf{0.33} & \textbf{0.30} & \textbf{0.33} \\
\midrule
\multicolumn{4}{c}{\textbf{Subjective Evaluation (MOS, 1--5)}} \\
\midrule
& on-screen only & on/off-screen & All \\
\midrule
MMAudio          & 2.24 & 2.15 & 2.19 \\
VGGSound original     & 2.97 & 2.97 & 2.97 \\
StereoFoley-base & 3.05 & 2.82 & 2.93 \\
StereoFoley-obj  & \textbf{3.54} & \textbf{3.37} & \textbf{3.46} \\
\bottomrule
\end{tabular}
\end{adjustbox}
\vspace{-0.3cm}
\end{table}

\subsection{Datasets}
For our StereoFoley-base experiments, we use VGGSound~\cite{chen2020vggsound}, a widely used public dataset for V2A and Foley sound research, containing approximately 200K video examples. Although the dataset nominally provides stereo audio, our analysis revealed that about $27\%$ of videos are effectively mono, with left and right channels nearly identical (average channel difference $<10^{-5}$). 
Acknowledging this limitation for stereo learning, we nevertheless used VGGSound as the primary training set for the StereoFoley-base model. Additionally, to also demonstrate our model’s potential when trained on a bigger dataset, we trained our base model on a mix of VGGSound and AudioSet~\cite{gemmeke2017audio}, a large-scale dataset of about 2M clips. We filtered AudioSet to remove overlap with VGGSound and excluded clips containing speech and music, making the subset more suitable for Foley sound generation research.

For StereoFoley-obj, we observed that object-aware stereo is underrepresented in VGGSound. Using our synthetic pipeline’s object tracker, we analyzed trajectories and sizes to identify suitable samples. We retained videos with no more than three scenes, at least one of which contained a single trackable object that was either stationary off-center or moved beyond $15\%$ from the screen center. We further excluded cases where objects were too small ($\leq2\%$ of the frame) or too large ($\geq60\%$). We found that only about $18\%$ of the dataset satisfied these criteria, and we refer to this subset as VGG-obj. Based on our own inspection of VGGSound’s original audio, we estimate that less than $10\%$ of these clips (roughly $1.8\%$ of VGGSound overall) exhibit clear object-aware stereo spatialization, making VGGSound unsuitable for stereo spatialization learning on its own. 
We generated stereo object-aware audio for the VGG-obj subset. Replacing only this subset, we fine-tuned StereoFoley-base into StereoFoley-obj while retaining natural audio ambience.

\vspace{-0.2cm}
\subsection{Experiment Setup and Baselines}

We used $T = 9.56$ second audio-visual segments from VGGSound, with audio resampled to $f_s=48$ kHz stereo. We standardized \( x_{\text{video}}\) video streams to $f_v=25$ fps at a resolution of $ H \times W = 224 \times 224$ to match Synchformer input requirements. Our stereo codec encoder produced audio embeddings of $T_z \times D_z = 224 \times 256$, and Synchformer with overlapping window produced video embeddings of $T_v \times D_v = 224 \times 768$. Our generative base DiT model has 24 attention layers, 256 channels, and an embedding dimension of 1536, totaling approximately $1.1$B parameters. For StereoFoley-base, we trained the model from scratch for up to $800$ epochs (about one week) with a global batch size of $512$ on 8×NVIDIA A100 GPUs, using AdamW with a learning rate of $10^{-4}$, weight decay of $10^{-3}$, and $2500$ warmup steps. For stage two, StereoFoley-obj model, we fine-tuned the base model for an additional 150 epochs on a modified VGGSound dataset with replaced VGG-obj videos. During training, we employed classifier-free guidance (CFG)~\cite{ho2022classifierfree} with dropout of $0.1$, and alternated conditioning between audio and text CLAP embeddings, drawing each modality with probability $0.5$. For inference, we used $100$ denoising steps with a CFG scale of $6.0$.

As baselines, we compare our system against MMAudio~\cite{Cheng2025mmaudio} and Kling-Foley~\cite{wang2025klingfoley}.We use the large variant of MMAudio with 44.1,kHz, mono output and $\sim$1.03B parameters. Kling-Foley is stereo systems with $\sim$1.5B parameters, making both models comparable in scale with ours. Being current state of the art in temporal alignment, both systems use similar Synchformer with RoPE method for synchronization. Despite producing stereo output, Kling-Foley neither discusses object-awareness nor reports any related or stereo-specific metrics. Also, as the authors did not release code or checkpoints, we include the results reported in their paper on the VGGSound test set.

\subsection{Evaluation Metrics}  
\label{metrics}

For objective evaluation, we follow the protocol of~\cite{Cheng2025mmaudio}. Specifically, we assess distributional similarity using Fréchet Distance (FD) and Kullback–Leibler (KL) divergence with PANNs~\cite{kong2020panns}, VGGish~\cite{gemmeke2017audio}, and PaSST~\cite{koutini2022efficient} backbones; audio quality with Inception Score (IS)~\cite{salimans2016improved}; semantic consistency with IB-score~\cite{viertola2024temporally}; and temporal alignment with DeSync~\cite{Cheng2025mmaudio}. We acknowledge that that these metrics may not not be fully suitable for high–sample-rate stereo sound evaluation: The backbones used for FD and KL (PANNs and PaSST at 32 kHz; VGGish at 16 kHz) operate at lower sampling rates and are defined for mono audio, meaning stereo outputs must be resampled and downmixed to mono, making the metrics largely insensitive to high-frequency content and stereo characteristics. However, given that our main contribution lies in stereo object-awareness, we treat these established metrics as adequate proxies for comparability with prior work. To capture stereo characteristics more directly, we additionally report Stereo-Score, calculated as the average absolute inter-channel difference normalized by power.

As no existing spatial AV alignment metrics are directly applicable to open-domain stereo object correspondence, we introduce a bin-alignment score (BAS) to evaluate the alignment between object trajectories and audio spatialization.
In videos from VGG-obj test, we compare the normalized horizontal object positions $\bar{w}[k]$ (Sec.~\ref{sec:synthdatapipeline}) with the normalized audio energy center of mass positions $\bar{a}[k]$, both quantized into three bins: left, center ($\pm5\%$), and right. The score is defined as:
$BAS = \tfrac{1}{K} \sum_{k=1}^K \mathbf{1}[\mathrm{bin}(\bar{w}[k]) = \mathrm{bin}(\bar{a}[k])]$, effectively showing a fraction of frames where the two bins match.
We report BAS separately for both object on-screen and object off-screen segments, as well as the combined score, reflecting both object awareness and permanence.

We complement the BAS score with a subjective evaluation. We randomly selected 10 videos from VGGS-obj test set: 5 where the object remained on-screen throughout, and 5 where the object enters or leaves the screen. We compared four systems: VGGSound ground truth audio, MMAudio, StereoFoley-base, and StereoFoley-obj. A user study was conducted with $180$ participants, who rated how well the sound aligned with the visual object on a 5-point scale (1 = poor, 5 = excellent). As most non-expert raters are unfamiliar with explicitly judging stereo consistency, we added hidden qualification clips with mono sound and left-right flipped sounds, disqualifying raters who rated them “good” or “excellent.”

\vspace{-0.2cm}
\subsection{Results} 
\label{objective_eval}

Table~\ref{tab:quant_comparison} compares our models with baselines on the VGGSound test set. Overall, StereoFoley performs on par with both baselines, with small advantages in some metrics and slight deficits in others, all three representing the current state-of-the-art in V2A generation. As noted in Section~\ref{metrics}, comparisons are imperfect: models differ in output format and Kling-Foley does not report every metric. MMAudio, being mono, likely benefits from mono FD backbones and shows the least distances, while ours and Kling-Foley yield similar $\text{FD}_\text{PaSST}$. KL differences are negligible across models, while our model outperforms MMAudio in IS score. We show IB score close to Kling-Foley, both being slightly behind MMAudio.  
Interestingly, although the differences are marginal, our base model attains the best synchronization score despite using simple video and audio latent addition, unlike RoPE or ConvMLPs in prior works. Additionally, StereoFoley-base achieves a Stereo-Score of 0.21, with the object-aware variant reaching 0.24, confirming stronger stereo separation. Training on larger dataset including AudioSet provides only minor gains to our base model, while StereoFoley-Obj shows slight drops in all metics, likely because of substituting part of VGGSound with synthetic VGG-obj data.

We report stereo object alignment results in Table~\ref{tab:alignment}. The upper part shows BAS score on the VGG-obj test subset. As expected, MMAudio, which generates mono audio, shows the weakest alignment. StereoFoley-base performs on par with VGGSound original, effectively learning the stereo characteristics of the dataset, while StereoFoley-obj achieves the highest alignment in both on-screen and off-screen segments, improving the combined score from 0.23 to 0.33. This demonstrates that synthetic data helps learn spatial correspondence between visual objects and audio, highlighting a key finding of our study: stereo awareness can be learned with an end-to-end framework when provided with appropriate data.

The lower half of Table~\ref{tab:alignment} reports Mean Opinion Scores (MOS) from our human study ($N=1,341$ ratings from 131 raters, 49 disqualified). Results are shown separately for on-screen only and for cases where object enters or exits screen. Consistent with BAS results, MMAudio scored lowest, StereoFoley-base matched VGGSound original audio ratings, and StereoFoley-obj achieved the highest scores. A Kruskal–Wallis test followed by Bonferroni-corrected pairwise comparisons confirmed that StereoFoley-obj significantly outperforms all other systems ($p<0.001$), MMAudio underperforms ($p<0.001$), while StereoFoley-base and VGGSound original are statistically indistinguishable ($p=0.80$). No significant effect was found between on-screen only and on/off-screen conditions ($p>0.05$). These results demonstrate that object-aware training yields perceptually meaningful benefits, and that our BAS metric aligns well with human judgment.

\section{Conclusion}

We presented an end-to-end framework for video-to-audio generation. 
We showed that our model achieves performance on par with state-of-the-art V2A models for full-band stereo audio generation, using a Synchformer video encoder with a simple latent-space matching strategy.
More importantly, we demonstrated that the main challenge for object-aware stereo audio generation is not architectural but data-related. By synthesizing object-aware stereo datasets, the same model can be trained to produce spatially accurate, object-position–aware audio. While our pipeline uses amplitude panning, it could be adapted to alternative mixing conventions, extended to Ambisonics using object trajectories, or scaled to multichannel video-to-audio generation. Future work will explore these directions, alongside improving dataset quality and increasing diversity with both real and synthetic examples.

\bibliographystyle{IEEEbib}
\bibliography{strings,refs}

\end{document}